\documentclass[pra,twocolumn,groupedaddress,floatfix,showpacs]{revtex4}

\usepackage{epsfig}
\usepackage{amsmath}
\usepackage{bm}

\begin{document}
\title{ Security of quantum cryptography using balanced homodyne detection}
\author{Ryo Namiki}
\email[Electronic address: ]{ namiki@qo.phys.gakushuin.ac.jp} 
\author{Takuya Hirano }
\affiliation{Department of Physics, Gakushuin University,
Mejiro 1-5-1, Toshima-ku, Tokyo 171-8588, Japan}
\date{December 12, 2002}
\begin{abstract}
In this paper we investigate the security of a quantum cryptographic scheme which utilizes balanced homodyne detection and weak coherent pulses (WCP). The performance of the system is mainly characterized by the intensity of the WCP and postselected threshold. Two of the simplest intercept and resend eavesdropping attacks are analyzed. The secure key gain for a given loss is also discussed in terms of the pulse intensity and threshold.\end{abstract}

%\keywords{}
\maketitle

%%%%%%%%%%%%%%%%%%%%%   part 1 %%%%%%%%%%%%%%%%%%%%%%%%%%%%%%%%%%%%%%%%%%%%%%%%
 \section{ Introduction}
%%%%%%%%%%%%%%%%%%
Quantum cryptography allows two parties, Alice (the sender) and Bob (the receiver), to share a random bit sequence, called key, which is unknown to the eavesdropper Eve \cite{rmp74}.

In the past years several quantum cryptographic schemes, which are mainly based on two-level quantum systems, have been proposed \cite{rmp74,e91,B92,Huttener1995}, and various theoretical studies on the security have been published \cite{inf1,A54,A57,l3,l2,br1,bs-attack}. A conventional security measure of quantum cryptographic system is the secure key gain which represents the secure key bits gain per signal \cite{l3,l2}.

  Recently, continuous variable quantum cryptographic schemes have been proposed \cite{squeezed,epr,con,alcon,coherent,postsel}. The security of those schemes is based on the commutation relation between the quadrature amplitudes of light field. The quadrature measurement is implemented by balanced homodyne detection \cite{leonhardt}. Since balanced homodyne detection involves phase-shift operation and the most conventional signal source is coherent pulses, a certain combination of phase modulations and homodyne detection with a coherent pulse should provide the simplest continuous variable schemes from the experimental side. In this point an interesting scheme is the application of homodyning on the phase coding four-state protocol, and this scheme includes a postselection process which is deeply related to the system performance \cite{hirano}. The importance of the postselection is discussed in Ref. \cite{postsel}.

%%%%%%%%%%%%%%%%%%%%%%%%%%%%

In this paper we investigate the security of a quantum cryptographic scheme which uses balanced homodyne detection and weak coherent pulses \cite{hirano}.
In Sec. \ref{2} we review the principles of the scheme and introduce basic quantities to describe the system. 
In Sec. \ref{3} we consider two of the intercept and resend eavesdropping attacks. By explicitly writing the density operator of the signal, all the detectable disturbances can be seen. The change of the quadrature probability distribution and the bit error rate (BER) caused by the attacks are shown. In Sec. \ref{4} we consider a beam splitting attack and then show the secure key gain for a given loss.
%%%%%%%%%%%%%%%%%%%%%%  part 2 %%%%%%%%%%%%%%%%%%%%%%%%%%%%%%%%%%%%%%%%%%%%%
 \section{ quantum cryptography using balanced homodyne detection \label{2}}
%%%%%%%%%%%%%%%%%%%%%%  part 2 %%%%%%%%%%%%%%%%%%%%%%%%%%%%%%%%%%%%%%%%%%%%%
In this section we review the protocol \cite{hirano} and introduce a density operator description. Then we derive the formula for the BER in the absence of Eve. The protocol is similar to that of the interferometric quantum cryptography using four nonorthogonal states \cite{B92} except quadrature measurement is performed by balanced homodyne detection with a strong local oscillator field \cite{leonhardt}.
\subsection{Protocol and basic quantities}
%%%%%%%%%%

The protocol is as follows:
Alice randomly chooses one of the four coherent states $\left\{ |\alpha\rangle,|i \alpha\rangle,|-\alpha\rangle,|-i\alpha\rangle  \right\} $ with $ \alpha >0 $ and sends it to Bob. If Alice uses a pulsed light source, the coherent state is the eigenstate of the annihilation operator $\hat{a}$ of the pulse mode \cite{kumar,loudon90}. Then Bob randomly measures one of the two quadratures $\{ \hat{x}_1,\hat{x}_2  \}$. We define the quadratrures by the relation $ \hat{a}= \hat{x}_1+ i \hat{x}_2$, thus $ [ \hat{x}_1,\hat{x}_2  ] = i/2 $.
After the transmission of a large number of pulses, Alice and Bob communicate through a classical channel and they divide the resulted data into two parts:\textit{correct-basis} data and \textit{wrong-basis} data. We say a pulse is correct-basis in the cases that Bob measures $\hat{x}_1$ when Alice sent $|\pm \alpha\rangle $ and Bob measures $\hat{x}_2$ when Alice sent $|\pm i\alpha\rangle $. The other cases, the pulse is called wrong-basis. For the correct-basis pulses Bob sets the threshold $x_0 (\ge 0)$ and constructs his bit sequence by the following decision:
\begin{eqnarray}
({\text{bit  value}}) = \left\{ 
	\begin{array}{ll}
           1  & {\text{if }}  x  >  x_0  \\
            0  &{\text{if }}     x   <-x_0\\
          inconclusive &  {\text{ otherwise,}}
\end{array} \right.  \label{deceq}
\end{eqnarray} where $x$ is the result of Bob's measurement.
Alice's bit values are determined by the different manner: Alice regards $\{ |\alpha \rangle ,|i\alpha \rangle \}$ as ``1'' and $\{ |-\alpha\rangle,|-i\alpha\rangle \}$ as ``0''.

The density operator of the signal sent by Alice is described by 
\begin{equation}
\hat{\rho} = \frac{1 }{ 4 } \left( | \alpha \rangle \langle \alpha |+| -\alpha \rangle\langle -\alpha |+| i\alpha \rangle \langle i\alpha |+| -i\alpha \rangle\langle -i\alpha | \right),
\end{equation}where the factor $\textstyle\frac{1 }{ 4 }$ denotes that each of the four states appears with equal probability. Eve's task is to distinguish the four states. Since the four states are not orthogonal with each other, complete differentiation is impossible.  
 If Alice announces the basis, i.e., the quadrature on which she encoded the bit information, then the density operator is reduced to 
 \begin{eqnarray}
\hat{\rho}_1&=&\frac{1}{ 2} \left( | \alpha \rangle \langle \alpha |+| -\alpha \rangle\langle -\alpha | \right) \label{dmr}
\end{eqnarray} or
\begin{eqnarray}
\hat{\rho}_2&=&\frac{1}{ 2}\left(  | i\alpha \rangle \langle i\alpha |+| -i\alpha \rangle\langle -i\alpha |\right) \label{dmi}
\end{eqnarray} for the announced quadrature $\hat{x}_1$ and $\hat{x}_2$, respectively.
 Thus Bob's task is in a sense to differentiate between the two states $ \{ |\alpha\rangle ,| - \alpha \rangle \}$.  The decision (\ref{deceq}) is a practical and efficient implementation of this task. Although various differentiation tasks have been studied based on generalized quantum measurement process \cite{Ban1,Sasaki1,Hutt2,A57}, in this paper we consider only the measurements which can be realized by conventional optics and detectors. 

For a quantitative description, it is convenient to introduce the probability density that the outcome $x_\phi $ is obtained by measuring $\hat{x} _\phi= \hat{x}_1\cos{\phi }+\hat{x}_2 \sin{\phi} $ of a coherent state $| \alpha \rangle$    
{\begin{eqnarray}
\left| \langle x_\phi |\alpha \rangle \right|^2 = %| \langle x_1 |\alpha {\textrm{ e}}^{-i\phi} \rangle |^2  \Big{|}_{x_1=x_\phi}\\
\sqrt{\frac {2 }{\pi}} \exp{\left[ -2 ( x_\phi-\alpha \cos{\phi} ) ^2\right] }.
\end{eqnarray}Then the probability distribution of quadrature measured by Bob is written as
\begin{widetext}
\begin{equation}
\langle x_i | \hat{\rho}_j |x_i \rangle 
= \left\{ 
	\begin{array}{lll}
           \frac{1}{ \sqrt{  2\pi}}\left\{ \exp{ \left[ -2 (x_i-\alpha ) ^2\right] }+\exp{\left[ -2 (x_i+\alpha ) ^2 \right] } \right\} \label{pd}  &{\text{ if}}  &  i=j \\
\\
     \sqrt{\frac{2}{ \pi}} \exp{ \left( -2  x_i  ^2 \right) }& {\text{ if}} & i \ne j ,
       \end{array} \right.  
\end{equation}  \end{widetext} with $i, j=1,2  $ (see Fig. \ref{distribution0}). $i=j$ is for correct-basis pulses and $i \ne j$ is for wrong-basis ones. If Alice announces the states she sent, Bob observes the quadrature distributions for the coherent states (see Fig. \ref{distribution0}). The quadrature distributions represent the conditional probability that characterizes the signal detection and thus any detectable disturbance should appear on the distributions. 
\begin{figure}[]
\epsfig{file=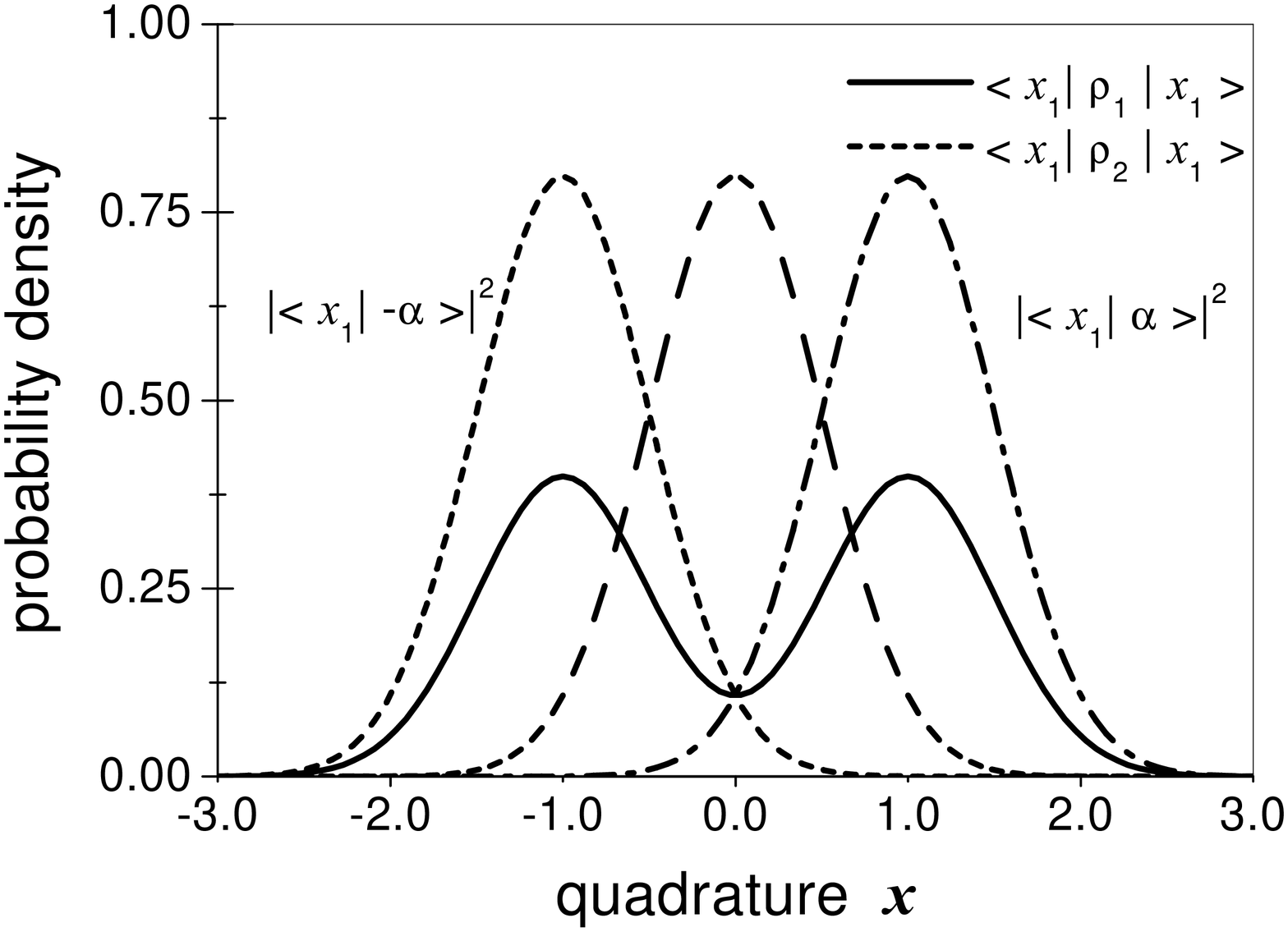,width=8.6cm} %qpvsx
 \caption{\label{distribution0}The quadrature distributions for correct-basis $\langle x_1|\hat{\rho}_1 | x_1 \rangle$, wrong-basis $\langle x_1|\hat{\rho}_2 | x_1 \rangle$ pulses and the coherent states $|\langle x_1| \pm \alpha \rangle|^2$ for the pulse intensity $n=\alpha ^2=1$ in the absence of the eavesdropper. The correct-basis and wrong-basis distributions are observed after Alice's announcement of the basis. In contrast to this, the distributions for the coherent states cannot be observed until Alice announces the states, and the announcement of the states sacrifices the key bits. 
The bit errors originate from the overlap between the Gaussian distributions $|\langle x_1 | \pm \alpha \rangle |^2$ and they are mainly distributed around $x=0$. Bob can efficiently cut out some of these bit errors by introducing the threshold $x_0$ and discarding the inconclusive results ( $|x| \le x_0$) when he constructs the key bits.}    
\end{figure}  

Since we can treat the quadratures $\hat{x}_1$ and $\hat{x}_2$ equally, hereafter we consider only the cases that Bob measures $\hat{x}_1$. Then we refer $|\pm \alpha \rangle $ as correct-basis pulses and $ |\pm i\alpha \rangle $ as wrong-basis pulses.

%%%%%%%%%%%%%%%%%%%%%%  part 3 %%%%%%%%%%%%%%%%%%%%%%%%%%%%%%%%%%%%%%%%%%%%%%%%
\subsection{ Postselection efficiency and bit error rate}
%%%%%%%%%%%%%%%%%%%%%%%%%
We define the {\textit{ postselection efficiency}} as the probability that the absolute value of correct-basis pulse's quadrature $|x|$ exceeds the threshold $x_0$. With this probability a correct-basis pulse gives a bit value according to decision (\ref{deceq}). The postselection efficiency in the absence of Eve is written as
\begin{widetext}
\begin{eqnarray}
\nonumber P(x_0,n)&=&\int_{-\infty}^{-x_0} \langle x_1 |\hat{\rho}_1 |x_1\rangle dx_1+\int_{x_0}^{\infty} \langle x_1 |\hat{\rho}_1 |x_1\rangle dx_1\\
&=& \frac{1}{2} \left\{  {\textrm{  erfc}}\left[  \sqrt{2}\left( x_0+\sqrt{n}\right) \right] + {\textrm{  erfc}}\left[  \sqrt{2}\left( x_0-\sqrt{n}\right) \right] \right\} \label{eff} ,
\end{eqnarray}\end{widetext}
where \begin{eqnarray}  {\textrm{ erfc}}(x)= \frac{2}{ {\sqrt{\pi } } }\int_x^\infty \exp{(-t^2)} dt, \end{eqnarray} and $n=\alpha^2 $ is the pulse intensity (the mean photon number per pulse).
If $x_0 =0 $, then every correct-basis pulse gives a bit value, thus $P(0,n)=1$ for any $n \ge 0$. For a given $x_0$ the BER can be written as the probability that Bob's measurement of $\hat{x}_1$ results an outcome $x_1< - x_0$ when Alice sent the state $|\alpha \rangle $ divided by $P(x_0,n)$ \begin{eqnarray}
 \nonumber q(x_0,n) &=& \frac{1}{ P(x_0,n)}   \int_{-\infty}^{-x_0}\left| \langle x_1|\alpha \rangle \right|^2 dx  \\ 
     &=&\frac{1}{2P(x_0,n)}{\textrm{ erfc}}\left[  \sqrt{2}\left( x_0+\sqrt{n}\right) \right] \label{ber}.\end{eqnarray}
$P$ and $q$ are decreasing functions of $x_0$. This means that Bob can obtain smaller BER sacrificing the efficiency by raising the threshold value.
 
Because the BER is a function of $n$ and $x_0$, the security of the system depends on \textit{a priori} selected $n$ and postselected $x_0$. The parameters should be determined so that the system provides higher security. For this purpose we consider the simplest eavesdropping attacks in the following sections.

%%%%%%%%%%%%%%%%%%%%%%  part  4 %%%%%%%%%%%%%%%%%%%%%%%%%%%%%%%%%%%%%%%%%%%%%%%
            \section{Eavesdropping: intercept and resend \label{3} }
In this section we consider two of the intercept and resend eavesdropping attacks. The density operators of the disturbed signals are explicitly shown and then the quadrature distributions and BERs are obtained.
%%%%%%%%%%%%%%%%%%%%%%%%%%%%%%%%%%%%%%%%%%%%%%%%%
          \subsection{Simultaneous measurement attack}
%%%%%%%%%%%%%%%%%%%%%%%%%%%%%%%%%%%%%%%%%%%%%%%%%

  Here we show the effects of a simultaneous measurement attack as a function of $n$. Eve's strategy is as follows. She splits the signal into two pulses of half intensity by using a 50:50 beam splitter (BS) and measures $ \hat{x}_1$ of one pulse and $\hat{x}_2$ of the other pulse. Then Eve obtains a pair of values $(\tilde{x}_1,\tilde{x}_2)$ for each signal. The inequality between $\tilde{x}_1$ and $\tilde{x}_2$ determines the most probable state as follows:
\begin{eqnarray} \left\{ 
	\begin{array}{llll}
     \ |\alpha \rangle &     &{\text{ if}} & \tilde{x}_1   \ge  |\tilde{x}_2 | \\
     |i \alpha \rangle &     &{\text{ if}} &\tilde{x}_2    >  |\tilde{x}_1 | \\
     |-\alpha \rangle  &     &{\text{ if}} &  -\tilde{x}_1    \ge  |\tilde{x}_2 | \\
     |-i \alpha \rangle &     &{\text{ if}} & -\tilde{x}_2     >  |\tilde{x}_1 | .
\end{array} \right.   \label{} \end{eqnarray} Thus Eve resends the signal to Bob according to this decision. It should be noted that the extra noise of simultaneous measurement is minimized in this measurement for the case of a coherent state \cite{leonhardt,gheizen}, however, this does not mean that this measurement gives the optimal differentiation of the four states. 

To write down the density operator of Eve's resending signal, let us consider the case that Alice sent $|\alpha \rangle $. The probability that Eve gets an outcome $(\tilde{x}_1,\tilde{x}_2)$ is given by the product of the two quadrature distributions for the split coherent states
 \begin{eqnarray}
\nonumber Q_{n}(\tilde{x}_1,\tilde{x}_2)&=& \left|  \Big{ \langle} x_1 \Big{|} \frac{\alpha }{ \sqrt{2}} \Big{ \rangle} \right| ^2  \left| \Big{ \langle } x_2 \Big{ |} \frac{\alpha }{ \sqrt{2}} \Big{ \rangle} \right| ^2\ \Bigg{ |}_{x_1=\tilde{x}_1, x_2=\tilde{x}_2} \\
&=&  \frac{ 2 }{ \pi} \exp{\left[ -2\left( \tilde{x}_1 -\sqrt{\frac{n}{2}} \right)^2-2\tilde{x}_2^2   \right] }. 
\end{eqnarray}
The resending signal can be characterized by three probabilities $p_+$, $p_\perp$, and $p_- $: 
The first one is that Eve resends the original state $| \alpha \rangle $ correctly
\begin{eqnarray}
p_+(n)&=&  \int_{x_1 \ge |x_2 |} Q_{n}(x_1,x_2)dx_1dx_2 .  
\end{eqnarray}
 The second one is that Eve resends either of $\pi /2 $-phase-shifted states $|\pm i \alpha \rangle $  \begin{eqnarray}
p_\perp (n) &=& \int_{x_2 > |x_1 |} Q_{n}(x_1,x_2)dx_1dx_2 .
\end{eqnarray}
The last one is the probability that Eve resends the $\pi $-phase-shifted state $|-\alpha \rangle $\begin{eqnarray}
p_-(n) &=& \int_{-x_1 \ge |x_2 |} Q_{n}(x_1,x_2)dx_1dx_2  .
\end{eqnarray}

Consequently, the original state is transformed as 
%\begin{widetext}
\begin{eqnarray}
|\alpha \rangle \langle \alpha| \to&\nonumber\\
& p_+|\alpha \rangle \langle \alpha |+p_-|-\alpha \rangle \langle -\alpha |\nonumber\\
&+p_\perp \left( |i\alpha \rangle \langle i\alpha | +|-i \alpha \rangle \langle -i \alpha | \right).
\end{eqnarray} The quadrature distribution of this signal for $n=1$ is shown in Fig. \ref{eb6}.
 %\end{widetext}
\begin{figure}%[]
\epsfig{file=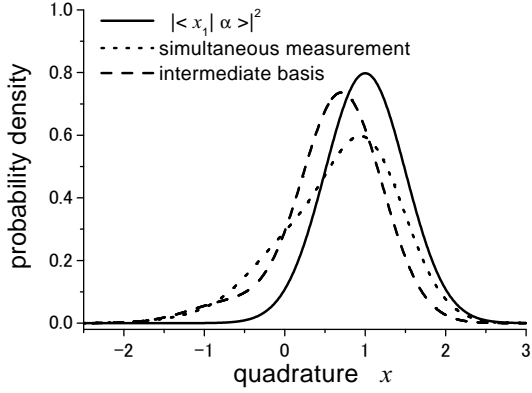,width=8.6cm}%eb6
\caption{\label{eb6} 
Disturbed quadrature distributions for the pulse intensity $n=1$. The solid line shows the quadrature distributions in the absence of Eve $|\langle x_1 | \alpha \rangle|^2 $.
The dotted line is for the simultaneous measurement attack, and the dashed line is for the intermediate basis attack. These distributions are observed if Alice announces the states.}
\end{figure} 

As a result, Bob's density operators after Alice's announcement of the basis, say $\hat{\rho}_1'$ and $\hat{\rho}_2'$, are expressed as
\begin{eqnarray}
\hat{\rho}_1'=(p_+ +p_-) \hat{\rho}_1+2p_\perp  \hat{\rho}_2, \\
\hat{\rho}_2'=2p_\perp  \hat{\rho}_1+(p_+ +p_- ) \hat{\rho}_2
.\end{eqnarray}
In these formulas the terms including $p_\perp$ indicate the part where Eve encodes the bit information on the wrong-basis. The quadrature distributions for the correct-basis pulse and wrong-basis pulse are given by $\langle x_1 |\hat{\rho}_1'| x_1\rangle $ and $\langle x_1 |\hat{\rho}_2'| x_1\rangle $, respectively (see Fig. \ref{eb5}). A remarkable point is that Bob observes the effects of eavesdropping in the wrong-basis data as well as in the correct-basis data.
\begin{figure}%[]
\epsfig{file=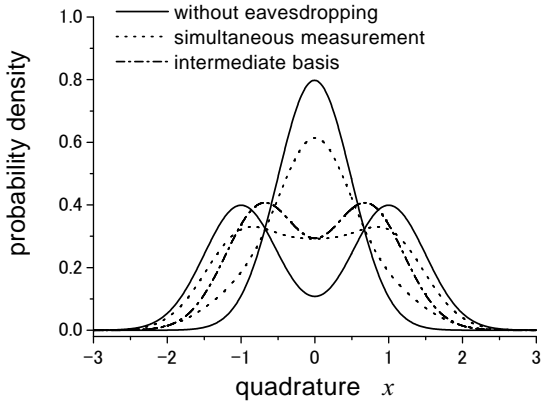,width=8.6cm}%eb5
\caption{  \label{eb5} The correct-basis distributions and wrong-basis ones for $n=1$. The solid lines are for the absence of Eve. The dotted lines describe the case that Eve performs the simultaneous measurement; the curve that has the maximum at $x=0$ is for wrong-basis pulses. The dash-dot line represents the result of the intermediate basis attack, where the correct-basis and wrong-basis distributions take the same form.}       
\end{figure}

To calculate Bob's BER in the presence of Eve, we rewrite the postselection efficiency 
\begin{eqnarray}
 \nonumber {P'}^{(1)}(x_0,n)&=&\int_{x_0}^{\infty} \langle x_1 |\hat{\rho}_1' |x_1\rangle dx_1+\int_{-\infty}^{-x_0} \langle x_1 |\hat{\rho}_1' |x_1\rangle dx_1\\
                    &=&(p_+ +p_-)P(x_0,n)+ 2p_\perp { \textrm{ erfc}}\bigl( \sqrt{2} x_0\bigr).\nonumber\\
\end{eqnarray}
With this equation we obtain Bob's BER   
\begin{widetext}
\begin{eqnarray}
q_{EB}^{(1)}(x_0,n)&=&\nonumber \frac{1}{{ {P'}^{(1)}(x_0,n)}}  \left( p_+\int_{x_0}^\infty \left|\langle x_1|-\alpha \rangle\right|^2dx +p_- \int_{x_0}^\infty \left| \langle x_1|\alpha \rangle\right|^2dx+2p_\perp\int_{x_0}^\infty \langle x_1|\hat{\rho}_2|x_1 \rangle   dx \right)\\ 
&=&\frac{1}{{ 2{P'}^{(1)}(x_0,n)}}  \left\{ p_+{\textrm{  erfc}}\left[ \sqrt{2} \left( x_0+\sqrt{n} \right) \right] +p_-{\textrm{  erfc}}\left[ \sqrt{2} \left(x_0-\sqrt{n} \right) \right] +2p_\perp { \textrm{ erfc}}\left(\sqrt{2} x_0\right) \right\}.\nonumber \\ 
\end{eqnarray}
\end{widetext}

After the classical communication Eve may know the basis on which Alice encoded the bit information for each pulse, and Eve's information is determined by the quadrature value of split signal measured in the ``correct-basis". Thus Eve's BER is $q_E^{(1)}(0,n)=q(0,n/2)$ (see Fig. \ref{bervsn}). Note that Eve's threshold is always zero because she cannot select the parts of the data that contribute to the key bits.

\begin{figure}%[hbtp]
\epsfig{file=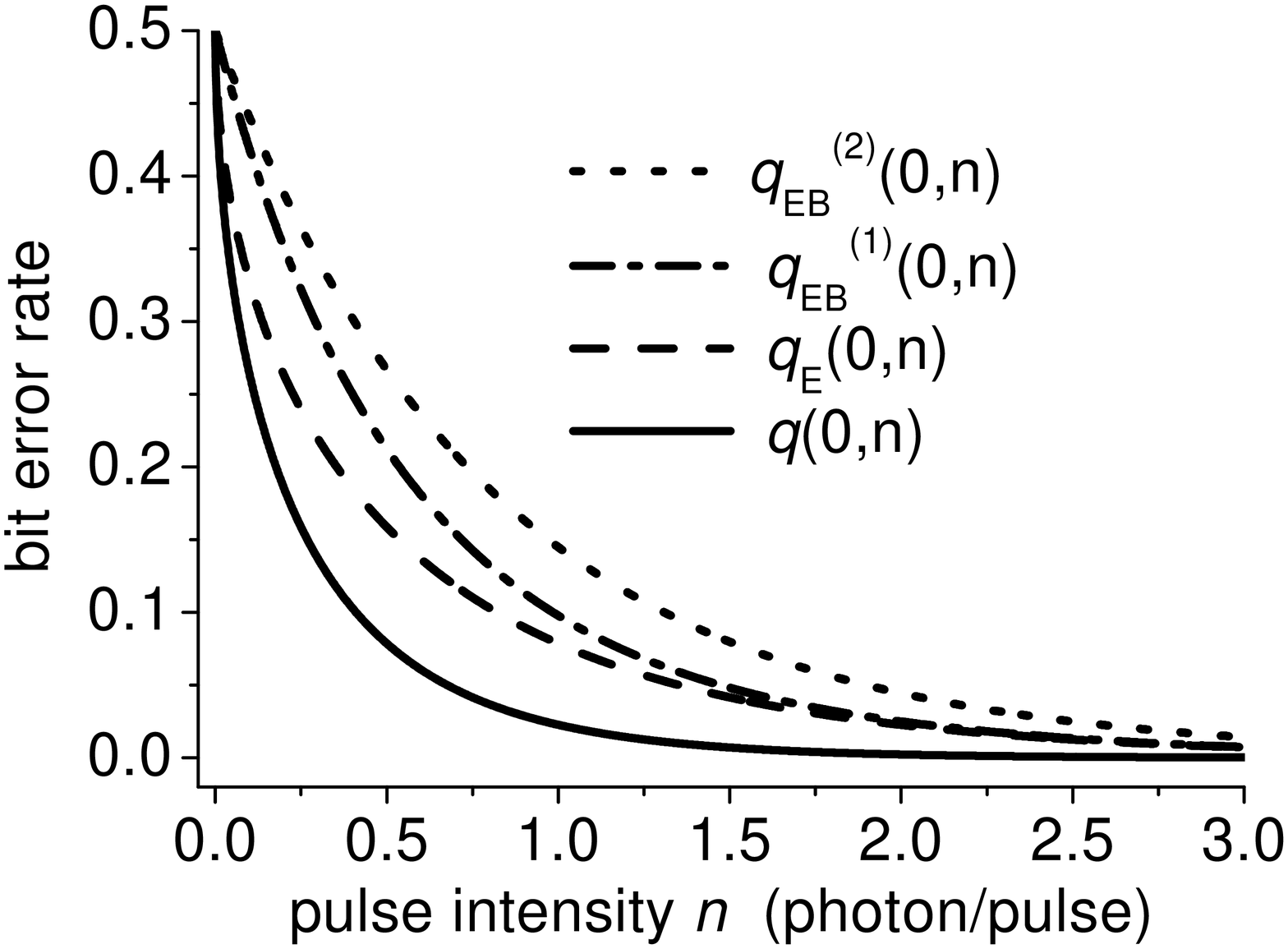,width=8.6cm}%bervsn
\caption{\label{bervsn} The BERs as functions of the pulse intensity $n$ for the threshold $x_0 =0$.
The solid line is Bob's BER in the absence of Eve $q(0,n)$. The dash-dot line shows Bob's BER for the simultaneous measurement attack $q_{EB}^{(1)}(0,n)$. The dashed line is for the intermediate basis attack $q_{EB}^{(2)}(0,n)$. The dotted line is Eve's BER for the two of the attacks $q_E^{(1)}(0,n)=q_E^{(2)}(0,n)=q (0,n/2)$.}
\label{gner}
\end{figure}

%%%%%%%%%%
%%%%%%%%%%%%%%%%%%%%%%%%%%%%%%%%%%%%%%%%%%%%%%%%%
  \subsection{Intermediate basis attack}
%%%%%%%%%%%%%%%%%%%%%%%%%%%%%%%%%%%%%%%%%%%%%%%%%
Here we consider an eavesdropping strategy that Eve measures the intermediate basis operator $\hat{x}_{\pi/4} \equiv \textstyle\frac{\hat{x}_1+\hat{x}_2}{\sqrt{2}}$. In the case that Eve performs single homodyne detection for each pulse, the intermediate basis minimizes her BER on the average.

By measuring $\hat{x}_{\pi/4}$, Eve obtains her BER $q_E^{(2)}(0,n)= q(0,n/2)$.
Since she has no information about Alice's choice of the basis, she encodes the bit information on the intermediate basis when she resends the pulse to Bob. Namely, Eve resends $|\alpha \exp(i\pi /4) \rangle $ if the outcome $x_{\pi/4} \ge 0 $ and $|-\alpha \exp(i\pi /4) \rangle $ otherwise. This operation transforms the four states as
\begin{eqnarray}
 &&|\pm \alpha \rangle\langle\pm \alpha | ,|\pm i\alpha \rangle \langle \pm i\alpha | \to \nonumber  \ \ \ \ \ \ \ \ \ \ \ \ \ \  \ \ \\ 
&&(1-q_E )| \pm \alpha e^{i\pi /4} \rangle \langle  \pm \alpha e^{i\pi /4}| + q_E | \mp\alpha e^{i\pi /4} \rangle \langle \mp\alpha e^{i\pi /4}|.\nonumber \\ \label{ad2}\end{eqnarray}
 An example of the resulted quadrature distributions is shown in Fig. \ref{eb6}. It yields the same distribution for the correct-basis pulse and wrong-basis pulse (see Fig. \ref{eb5}). This feature originates from the fact that Eve has no information about Alice's choice of the basis.

From transformation (\ref{ad2}), Bob's postselection efficiency is given by 
\begin{widetext}\begin{eqnarray}
 \nonumber {P'}^{(2)}(x_0,n) &=&\int_{x_0}^{\infty} \left( \left| \langle x_1 |\alpha e^{i \pi /4 }\rangle \right|^2+ \left|\langle x_1 |-\alpha e^{i \pi /4 }\rangle \right|^2 \right) dx_1 \\
&=& P(x_0 ,n/2).
\end{eqnarray}
Then, considering the case that Alice sends $|\alpha \rangle $ and Bob gets $x_1 <-x_0$ we obtain Bob's BER (see Fig. \ref{bervsn}):   
\begin{eqnarray}
 q_{EB}^{(2)}(x_0,n)&=&\nonumber \frac{1}{{P'}^{(2)}(x_0,n)}\int_{-\infty}^{-x_0} dx \left( \left[ 1-q_E(n) \right] \left| \langle x_1 | \alpha e^{i \pi /4 }\rangle \right|^2 +q_E(n) \left| \langle x_1 | -\alpha e^{i \pi /4 }\rangle \right|^2 \right) \\
&=&[1-q_E(n) ]q( x_0,n/2) + \frac{q_E(n)}{ 2 {P'}^{(2)}(x_0,n)} {\textrm{ erfc}}\left[ \sqrt{2}\left( x_0 - \sqrt{n/2}\right) \right] .
\end{eqnarray} \end{widetext}

%%%%%%%%%%%%%%%%%%%%%%%%%%%%%%%%%%%%%%%%%%%%%%%%%%%%%%%%%%%%
As shown in Fig. \ref{eb6}, the non-Gaussian distribution reveals the presence of Eve. This feature can also be seen in the wrong-basis distribution in Fig. \ref{eb5}. The monitoring of the wrong-basis distribution profile restricts Eve's operation. For instance, suppose that Eve resends a coherent state for each pulse. Then the density operator after Alice's announcement of the basis can be written as the mixture of coherent states \begin{eqnarray}
\hat{\rho}_2' = \sum _{r,\theta} w(r, \theta)| r e^{i\theta} \rangle \langle r e^{i\theta} | ,\label{cmix} \\
\text{with }\sum _{r,\theta} w(r, \theta)=1 ,\ w( r, \theta ) \ge 0, \nonumber\\ 
\nonumber \text{for } r\ge 0\text{ and } -\pi \le \theta < \pi.
\end{eqnarray} 
If Bob observes no disturbance in the wrong-basis distribution, i.e., $\langle x_1| \hat{\rho}_2'| x_1 \rangle=\langle x_1 | \hat{\rho}_2|x_1 \rangle$, we find the condition $w(r,\theta ) =0$ for all $ \theta \neq \pm\pi / 2$. The trivial solution which satisfies this condition is that Eve resends the vacuum state [$w(r,\theta ) =0$ for $r \neq 0 $] which induces Bob's BER of 0.5. The other solutions require that Eve distinguishes between $\hat{\rho}_1 $ and $\hat{\rho}_2$ without error.  
Similarly, the monitoring of the correct-basis distribution profile gives additional limitations on Eve's operation. 
 
 Although the security proof needs more analysis about Eve's operation and leaked information, most of the attacks cannot work provided Bob observes the distributions carefully.
An undetectable attack is the beam splitting attack shown in the next section.   
%%%%%%%%%%%%%%%%%%%%%%%%%%%%%%%%%%%%%%%%%%%%%%%%%%%%%%%%%%%%
\section{secure key gain for a given loss\label{4}}
%%%%%%%%%%%%%%%%%%%%%%%%%%%%%%%%%%%%%%%%%%%%%%%%%%%%%%%%%%%%
For any practical implementation of quantum cryptography, the transmission loss is an unavoidable problem \cite{br1,bs-attack}. The loss weakens the signal intensity and in the same time it potentially causes the information leakage to Eve. In order to estimate the performance of the system for a given loss, we consider a beam splitting attack and calculate the secure key gain which gives the secure bits gain per signal \cite{l3,l2}. The threshold and pulse intensity can be selected to maximize the secure key gain.
 
We assume that (1) Eve has a lossless optical fiber, and (2) Eve can store the light pulses for an arbitrarily long time before measuring them. Then Eve replaces the original transmission path with her lossless fiber and splits the pulse using an asymmetric BS with reflection efficiency $1-\eta $ which is equal to the original transmission loss, so that Eve does not change the pulse intensity received by Bob. Further, she measures her split signal in the correct-basis by delaying her measurements until she have received the basis information.

  After the error correction Alice and Bob share the same bit sequence, thus the question is how well are the outcomes of Eve's measurements correlated to the transmitted states. The correlation is estimated by the relative expected collision probability $ P_c$ which determines the fraction $\tau=1+\log_{2}{P_c} $, by which the bit sequence is shortened in the privacy amplification process \cite{gpa}. Finally, Alice and Bob can bound Eve's Shannon information about the key by discarding $s$ bits, \begin{eqnarray}
I_E \le \frac{2^{-s}}{ \ln 2} .
\end{eqnarray}
         
  Let us consider the signal transmission that the binary signals $\{ | \alpha \rangle , | -\alpha \rangle \}$ are sent and the homodyne detection $ \int dx_1 |x_1 \rangle\langle x_1|$ is performed. The collision probability for the beam splitting attack is determined as follows:
According to \cite{A54}, the relative expected collision probability can be written as
\begin{eqnarray}
%\langle P_c\rangle ^{1/N} &=& \sum_{k,\alpha,\Psi} \frac{P^2(\Psi_\alpha,k_\alpha%)}{ p(k_\alpha)}\\
 P_c &=& \sum_{\beta=\pm \alpha } \sum_{x_1} \frac{p^2 (x_1 ,\beta )}{ {\textrm{ Prob}} (x_1)} ,
\end{eqnarray}
where $p(x_1 ,\beta )$ is the joint probability of the sent signal $| \beta \rangle$ and the measured quadrature value $x_1$, and ${\textrm{ Prob}} (x_1)$ is the probability that the measurement results $x_1$,  \begin{eqnarray}
{\textrm{ Prob}}(x_1)=\langle x_1 |\hat{\rho}_1|x_1 \rangle .
\end{eqnarray}
 The conditional probability that the state is $|\alpha \rangle$ when the measurement results $x_1$ is given by   
\begin{eqnarray}
{\textrm{ Prob}}(\alpha |x_1)=\frac{\left| \langle x_1|\alpha \rangle \right|^2 }{ \left|\langle x_1|\alpha \rangle \right|^2+\left| \langle  x_1|-\alpha \rangle \right|^2 }.
\end{eqnarray}
Then we obtain the joint probability by the product $p(x_1,\alpha )={\textrm{ Prob}}(\alpha |x_1){\textrm{ Prob}}(x_1)$.
Thus the fraction is written as a function of the pulse intensity $n=\alpha  ^2$:
\begin{widetext}\begin{eqnarray}
\tau (n) &= &1+\log_2{P_c} \nonumber \\ 
    &=&1+\log_2 {\left( \sqrt{ \frac{2}{ \pi  }} \int_0^{\infty}dx \frac{   {\textrm{exp}}\left[ -4\left( x-\sqrt{n} \right) ^2\right]+ {\textrm{exp}}\left[ -4\left( x+\sqrt{n}\right)^2\right]  }{ \exp{ \left[ -2\left( x-\sqrt{n}\right)^2\right] } + \exp{ \left[ -2\left( x+\sqrt{n}\right)^2\right] } }   \right)}, 
\end{eqnarray}%\end{widetext}
where we replaced the $x_1$ summation to the integration $\int_{-\infty}^\infty  dx_1 $.

Using this expression we obtain the secure key gain (with ideal error correction) \cite{l2} 
%\begin{widetext}
\begin{eqnarray}
G(x_0,n,\eta)&=& \frac{1}{2 } P(x_0,\eta n) \left[  I_{AB}  { \bm ( }x_0,\eta n{ \bm ) } -\tau {\bm ( }(1-\eta)n{\bm )} \right],
\end{eqnarray}
where \begin{equation}
I_{AB}(x_0,n) = \sum_{|x|>x_0}\frac{\textrm{Prob}(x)}{P(x_0,n)} \left[ 1+ \textrm{Prob}(\sqrt{n} |x) \log _2{\textrm{Prob}(\sqrt{n} |x)} +\textrm{Prob}(-\sqrt{n} |x) \log _2{\textrm{Prob}(-\sqrt{n} |x)}\right]
\end{equation}\end{widetext}
 is the mutual information between Alice and Bob \cite{inf1,A57}, and  $(1-\eta)n$ is the lost pulse intensity received by Eve, and $\eta n$ is the pulse intensity received by Bob. 
The loss region where the secure key is obtainable is defined by $G>0$. The region is determined by the sign of $I_{AB}-\tau$ because $P$ is always positive. Since $I_{AB}$ is an increasing function of $x_0$ and $I_{AB} \to 1 (x_0 \to \infty)$, it is always possible to find a sufficiently large value of $x_0$ that satisfies $G> 0$ for any $\eta > 0$ and $n >0$  if $\tau <1$.   

Typical behavior of $G$ as a function of $1-\eta $ for a fixed $n$ is shown in Fig. \ref{g1}. For a given value of $x_0$, $G$ rapidly vanishes at a certain value of $1-\eta$. By raising the value of $x_0$ we can extend the loss region where the secure key is obtainable. For fixed $n$ and $1-\eta$, $P$ is a decreasing function of $x_0$ and $I_{AB}$ is an increasing function of $x_0$. Thus, the trade-off between $P$ and $I_{AB}$ determines the optimal threshold which maximizes $G$ [see Fig. \ref{s-rate}(a)]. Then by comparing $G$ for various $n$ at the optimal threshold, we can optimize $n$ and $x_0$ simultaneously to maximize $G$ for a given value of $1-\eta$. The optimal value is $G=0.27$ ($x_0= 0.22$,  $n=0.89 $) for 10\% loss, $G=4.0 \times 10^{-2} $ ($x_0= 0.64$, $n= 0.62 $) for 50\% loss, and $G=6.0 \times 10^{-6}$ ($x_0= 1.91$, $n=0.59 $) for 90\% loss.
Figure \ref{s-rate}(b) shows $G$ for several values of $n$ at the optimal threshold as a function of $1- \eta$. 
\begin{figure}%[hbtp]
\epsfig{file=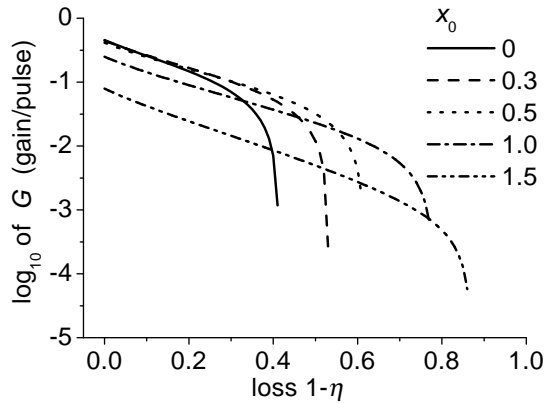,width=8.6cm}%gain-th2
\label{srate}
\caption{\label{g1} The secure key gain $G$ as a function of the loss $1-\eta$ for the pulse intensity $n=1$  and the threshold values $x_0 = 0, 0.3, 0.5, 1.0, 1.5 $.}
\end{figure}
\begin{figure}%[hbtp]
\epsfig{file=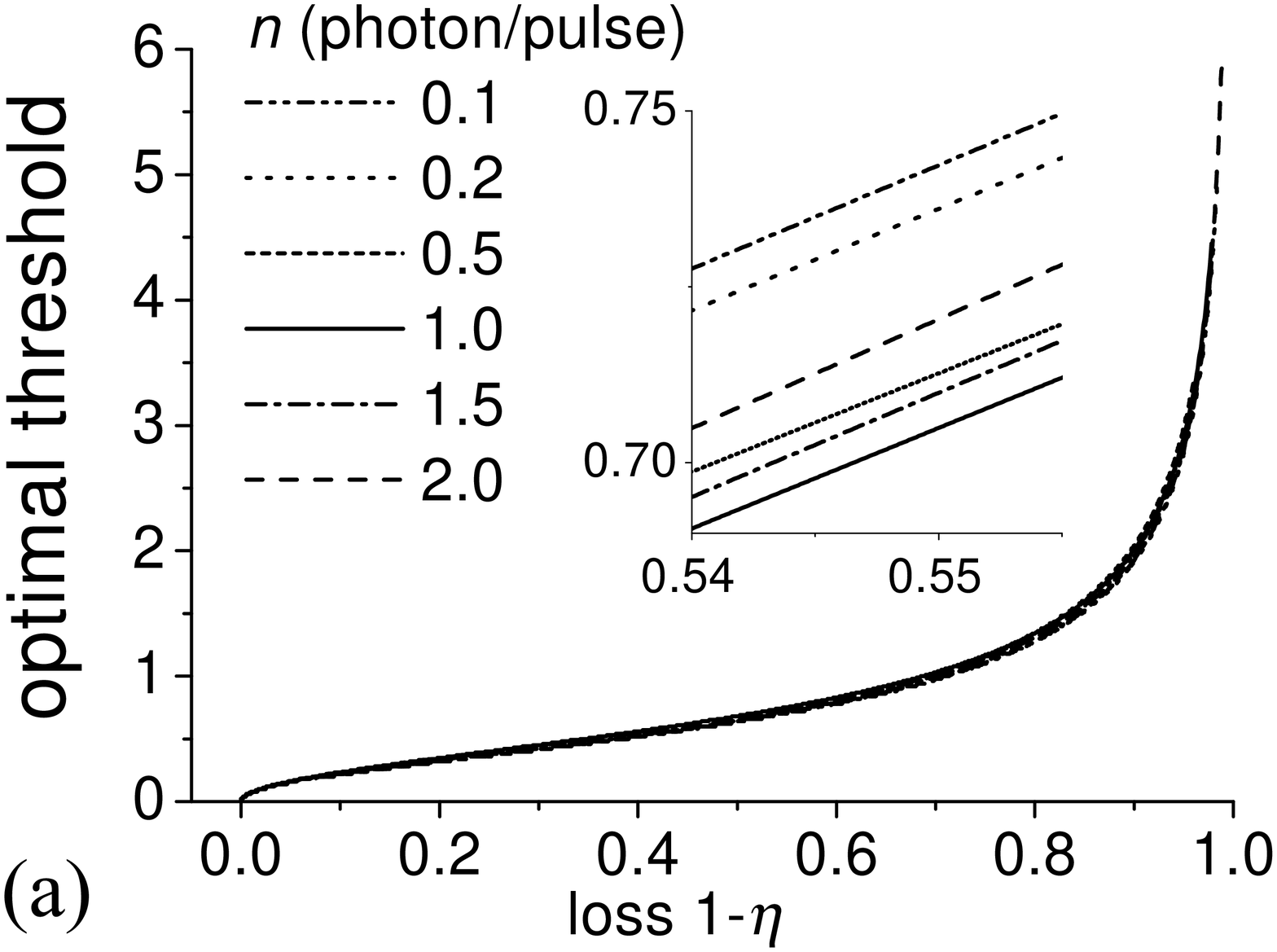,width=8.6cm}
\epsfig{file=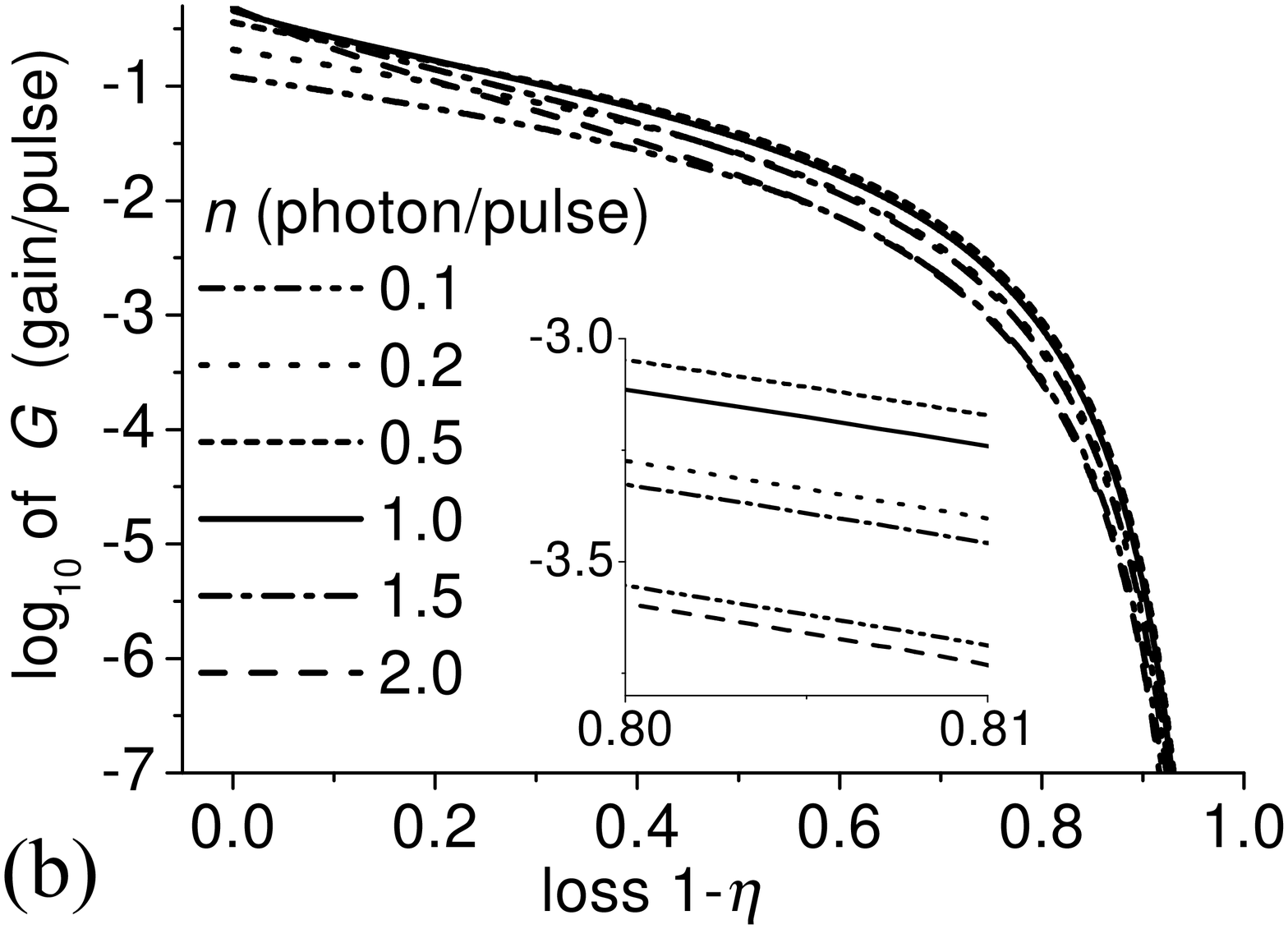,width=8.6cm}
\caption{\label{s-rate} (a) The optimal thresholds for the values of the pulse intensity $n=0.1, 0.2, 0.5, 1.0, 1.5, 2.0$ as functions of the loss $1-\eta$. (b) The key gain $G$ at the optimal threshold.}
\end{figure} 

The limitation for the key gain against the beam splitting attack under the assumption that Eve can use a positive operator valued measure for the split individual pulse is presented in Ref. \cite{namiki2}.

	%%%%%%%%%%%%%%%%%%%%%%%%%%%%%%%
\section{conclusions and remarks}
We investigated the security of quantum cryptography using balanced homodyne detection. For the two of the intercept and resend attacks, we derived the disturbed density operator of the signal and showed the quadrature distributions and BER as functions of the pulse intensity and postselected threshold. Both of the attacks are detectable only by the monitoring of the wrong-basis distribution profile. Our examples show that Eve's possible operations are greatly restricted if she does not disturb the quadrature distributions. However, in our formula, which describes only the signal density operator, the relation between the observed signal disturbance and potentially leaked information is unclear. Further analysis needs more general description.

We presented the secure key gain for a given loss, provided Eve performs a beam splitting attack. We can extend the loss region where the secure key is obtainable by raising the threshold value. The pulse intensity and threshold can be selected to maximize the secure key gain.

\begin{acknowledgments}
This work was supported by CREST, JST,
and ``R\&D on Quantum Commun. Tech."of MPHPT.
\end{acknowledgments}

%%%%%%%%%%%%%%%%%%%%%%%%%%%%%%%%%% reference

%\begin{widetext}


\begin{thebibliography}{}  %\label{sec:TeXbooks}
%\bibitem{Experimental Quantum Cryptography} C. H. Bennett, F. Bessette, G. Brassard, L. Salvail, and J. Smolin, J. Cryptology \textbf{5,} 3 (1992). 
\bibitem{rmp74}N. Gisin, G. Ribordy, W. Tittel, and H. Zbinden, \rmp \textbf{74,} 145 (2002)
%\bibitem{los}C. Marand and P. D. Townsend,  Opt.Lett.{\textbf{ 20,}}1695(1995).% "Quantum key distribution over distances as long as 30 km",


\bibitem{e91}A. K. Ekert, \prl\textbf{67,} 661 (1991).
\bibitem{B92} C. H. Bennett, \prl\textbf{68,} 3121 (1992).
\bibitem{Huttener1995}B. Huttner, N. Imoto, N. Gisin, and T. Mor, \pra\textbf{51,} 1863 (1995).
%\bibitem{conjugate1}S. J. D. Phoenix,  \pra\textbf{48,} 96 (1993).% "Quantum cryptography without conjugate coding",
%\bibitem{conjugate2}S. M. Barnett and S. J. D. Phoenix, \pra {\textbf{ 48,}} R5 (1993).% "Information-theoretic limits to quantum cryptography"
\bibitem{inf1} A. K. Ekert, B. Huttner, G. M. Palma, and A. Peres, \pra\textbf{50,} 1047 (1994).%"Eavesdropping on quantum-cryptographical systems",
%\bibitem{inc}H. E. Brandt, \pra\textbf{62,} 04231 (2000).

\bibitem{A54}N. L\"utkenhaus, \pra\textbf{54,} 97 (1996).%"Security against eavesdropping in quantum cryptography", 
\bibitem{A57} B. A. Slutsky, R. Rao, P. C. Sun, and Y. Fainman, \pra\textbf{57,} 2383 (1998).%"Security of quantum cryptography against individual attacks",
\bibitem{l3}N. L\"utkenhaus, \pra\textbf{59,} 3301 (1999).
\bibitem{l2}N. L\"utkenhaus, \pra\textbf{61,} 052304 (2000).
\bibitem{br1}G. Brassard, N. L\"utkenhaus, T. Mor, and B. C. Sanders, \prl\textbf{85,} 1330 (2000). % "Secrity Aspects of Practical Quantum Cryptography",
\bibitem{bs-attack} J. Calsamiglia, S. M. Barnett, and N. L\"utkenhaus, \pra \textbf{ 65,} 012312(2001).

\bibitem {squeezed} M. Hillery, \pra\textbf{61,} 022309 (2000).
\bibitem {epr} M. D. Reid, \pra\textbf{62,} 062308 (2000).
\bibitem {con} T. C. Ralph, \pra\textbf{62,} 062306 (2000).
\bibitem{alcon}N. J. Cerf, M. L\'evy, and G. Van Assche, \pra \textbf{63,} 052311 (2001).
\bibitem{coherent}F. Grosshans and P. Grangier, \prl\textbf{88,} 057902 (2002).
\bibitem{postsel} Ch. Silberhorn, T. C. Ralph, N. L\"utkenhaus, and G. Leuchs, \prl\textbf{89,} 167901 (2002).
\bibitem{leonhardt} U. Leonhardt, {\textit{Measuring the Quantum State of Light} } (Cambridge University Press, Cambridge, 1997).
\bibitem{hirano}T. Hirano, T. Konishi, and R. Namiki, \eprint{ quant-ph/0008037}.

\bibitem{kumar}P. Kumar, O. Ayt\"ur, and J. Huang, \prl\textbf{64,} 1015 (1990).
\bibitem{loudon90}K. J. Blow, R. Loudon, S. J. D. Phoenix, and T. J. Shepherd, \pra\textbf{42,} 4102 (1990).


\bibitem{Ban1}M. Ban, M. Osaki, and O. Hirota, \jmo\textbf{43,} 2337 (1996).
\bibitem{Sasaki1} M. Sasaki and O. Hirota, \pra\textbf{54,} 2728 (1996).
\bibitem{Hutt2}B. Huttner, A. Muller, J. D. Gautier, H. Zbinden, and N. Gisin, \pra\textbf{54,} 3783 (1996).
\bibitem{gheizen}E. Arthurs and M. S. Goodman, \prl\textbf{60,} 2447 (1988).%"Quantum Correlations: A Ganaralized Heisenberg Uncertainty Relation",


%\bibitem{oht}D. T. Smithey, M. Beck, M. G. Raymer, and F. Faridani, \prl\textbf{70,} 1244 (1993). 
%\bibitem{crec}G. V. Assche, J. Cardinal, and N. J. erf, E-print cs.CR/010730.
\bibitem{gpa}C. H. Bennett, G. Brassard, C. Cr\'epeau, and U. M. Maurer, IEEE Trans. Inf. Theory, \textbf{41,} 1915 (1995).%"Genelalized Privacy Amplification",

\bibitem{namiki2}R. Namiki and T. Hirano, (unpublished).
\end{thebibliography}
\end{document}